# Absence of orbital current torque in Ta/ferromagnet bilayers

Qianbiao Liu,[1] Lijun Zhu[1,2]*

[1] State Key Laboratory of Semiconductor Physics and Chip Technologies, Institute of Semiconductors, Chinese Academy of Sciences, Beijing 100083, China

[2] Center of Materials Science and Optoelectronics Engineering, University of Chinese Academy of Sciences, Beijing 100049, China

*ljzhu@semi.ac.cn

**Abstract:** It has become a heated debate as to whether the orbital Hall effect of a material could generate a non-local orbital current and a non-zero spin-orbit torque on an adjacent magnetic layer. Here, we report unambiguous evidence that, regardless of the ferromagnets (FMs) (e.g., Ni, $Ni_{81}Fe_{19}$, Fe, $Fe_{60}Co_{20}B_{20}$, and $Fe_{50}Pt_{50}$), the spin-orbit torque generated by an adjacent Ta, which is predicted to have a 50 times greater positive orbital Hall conductivity than the negative spin Hall conductivity, has essentially the same, negative efficiency. This agrees with the spin Hall effect of Ta being the only source of the interfacial torque. We identify that the constant, positive estimate of the torque of the Ta/FM samples from spin-torque ferromagnetic resonance (ST-FMR) analysis in a specific FM thickness range (≥2 nm for Ni) results from the overlook of a significant thick-dependent self-induced ST-FMR signal of the FM. These results indicate the absence of orbital current torque in Ta/ferromagnet systems, regardless of the type and the layer thickness of the ferromagnets.

## Introduction

The development of fast, energy-efficient memory and computing technologies has triggered bloomed interest in the spin-orbit torques (SOTs) exerted on a ferromagnet (FM) by spin currents.[1-3] Since the discovery of the SOTs,[1,2] it had been a consensus that the spin currents associated with the SOTs are only *directly* generated by the spin Hall effect (SHE)[1,3-8] or interfacial spin-orbit coupling (SOC) effects[2,9-14] and that the orbital angular momentum is highly localized and gets completely quenched within a very short length scale of 0.2-0.4 nm, regardless of the SOC.[15-17] Until very recently, some experiments,[18-23] which cannot be readily understood by directly generated spin currents, together with some orbital current theories,[24,25] stimulated searching for *indirect* sources of spin currents from orbital-to-spin conversion via a bulk or interfacial SOC (Fig. 1a).

Searching for hints for indirectly generating spin currents from the flow of orbital angular momentum has now become a rapidly growing interest in the field of spintronics.[26-40] For the sake of convenience, below we use orbital current torque to rename the SOTs with an orbital current as the initial source of the spin current. However, to be precise, experiments cannot directly tell whether the measured SOTs are initially due to a spin or orbital current because orbital current cannot interact with magnetization until being converted into a spin current.

So far, the most likely observations that support the existence of orbital torque rely on the use of Ni as the torque detector.[20,22,32-35] The dampinglike torque efficiency ($\xi_{DL}^j$) of Ta/Ni bilayers was analyzed, e.g., using spin-torque ferromagnetic resonance (ST-FMR), as positive in the Ni thickness range of a few nanometers, in striking contrast to the negative spin Hall ratio of Ta as measured using bilayers of Ta/non-Ni 3$d$ FMs (e.g., $Ni_{81}Fe_{19}$, Fe, FeCoB).[20,22] The argument from this observation in the literature was usually that Ni had a weak but slightly stronger SOC than $Ni_{81}Fe_{19}$, Co, Fe, and $Fe_{60}Co_{20}B_{20}$ such that Ni more efficiently converted the orbital current from Ta into spin current and generated a much stronger positive orbital torque than the negative spin Hall torque on the FM (the orbital Hall conductivity of Ta is predicted by band structure calculation to be 20-50 times greater than and of opposite sign compared to its spin Hall conductivity[41,42]). To explain the different experiments, such orbital torque argument had to assume that the orbital relaxation length ($\lambda$) of the sputter-deposited polycrystalline Ni was less than 0.5 nm (as indicated by the constant SOT for Ni thicker than < 3 nm[20], assuming the orbital current decays exponentially with the travel distance $t$ in analog to spin current[43,44], $j_L \sim e^{-t/\lambda}$, when the thickness of the layer in which the orbital current diffuses is greater than $\lambda$, Supplementary Fig. S1) or greater than 4 nm (as indicated by the continuous increase of the SOT with Ni thickness up to 20 nm[32-34]). Following similar arguments and with Ni as the detector, the generation of orbital current torque has been also claimed for many light metals that were believed to have little contribution to SOT generation before, such as Cr [19,26,29] and Ti [32,36]. Given the widespread impact and the striking disagreements between the orbital-current-torque arguments and the long-standing consensuses and between different orbital torque arguments, a careful experimental test of the in-depth physics and a unified, precise understanding of the Ta/ferromagnet are urgently required.

Here, we demonstrate evidence for the absence of orbital current torque in the prototype Ta/FM systems, regardless of the type and the layer thickness of the FM. We find that the positive $\xi_{DL}^j$ of the Ta/Ni, which is heavily cited in the literature as the most robust evidence of orbital current torque, is a misinterpretation of a non-negligible self-induced ST-FMR signal of the FM layer.

## Results

**Sample preparation**. We sputter-deposited in-plane magnetized Ta/FM bilayers with Ta thickness of 5nm and with the FM of Ni, $Ni_{81}Fe_{19}$, $Fe_{60}Co_{20}B_{20}$, Fe, and $Fe_{50}Pt_{50}$ and the control FM single-layer samples to study the SOTs generated by the SHE and the orbital Hall effect of Ta as a function of the type, the thickness ($t_{FM}$), and the SOC strength of the FM and the interface. Each bilayer is grown on an oxidized silicon substrate and protected by a MgO 2/Ta 2 bilayer (numbers are the layer thicknesses in nm) that is fully oxidized upon exposure to the atmosphere.[45] These samples are patterned into ST-FMR microstrips by photolithography and ion milling, followed by deposition of contacts of Ti 5/Pt 150. The resistivity of the Ta layer is determined to be 200 μΩ cm.

**ST-FMR measurement of interfacial spin-orbit torques.** We measure the SOTs of the Ta/FM using the three-terminal ST-FMR technique using the longitudinal voltage response (via magnetoresistance effects) by sweeping the in-plane magnetic field ($H$) at the azimuth angle ($\varphi$) of 45° relative to the rf current direction (Fig. 1b). As shown in Fig. 1c, each of the devices exhibits strong ST-FMR response ($V_{mix}$), fit of which to the relation[46]

$$V_{mix} = S \frac{\Delta H^2}{\Delta H^2 + (H-H_r)^2} + A \frac{\Delta H(H-H_r)}{\Delta H^2 + (H-H_r)^2}, \quad (1)$$

yields the magnitudes of the symmetric and antisymmetric components ($S$ and $A$), the FMR linewidth ($\Delta H$), and the resonance field ($H_r$). Since the $\sin2\varphi\cos\varphi$ scaling of $S$ and $A$ (Supplementary Fig. S2) have revealed no indication of any perpendicular spins[47] or perpendicular Oersted field [48], we define the ST-FMR efficiency ($\xi_{FMR}$) as

$$\xi_{FMR} = \frac{S}{A}\frac{e\mu_0 M_s t_{FM} d_{HM}}{\hbar}\sqrt{1+M_{eff}/H_r}, \quad (2)$$

where $e$ is the elementary charge, $\mu_0$ the permeability of the vacuum, $M_s$ the saturation magnetization of the FM, and $\hbar$ the reduced Planck's constant. $M_{eff}$ is the effective demagnetization field of the FM and can be determined from the rf frequency ($f$) dependence of $H_r$ following the Kittel's equation (Supplementary Fig. S3). As shown in Fig. 1d, $\xi_{DL}^j$ and the fieldlike torque efficiency ($\xi_{FL}^j$) can be estimated from the inverse intercept of the linear fit of $1/\xi_{FMR}$ vs $1/t_{FM}$ in the small-thickness regime following[5]

$$\frac{1}{\xi_{FMR}} = \frac{1}{\xi_{DL}^j}(1 + \frac{\hbar\xi_{FL}^j}{e\mu_0 M_s t_{FM} d_{HM}}). \quad (3)$$

So far, we have assumed that in the linear regime $S$ and $A$ included only negligible "artifacts" contributions ($S_{art}$, $A_{art}$) from spin pumping ($S_{art}\neq0$, $A_{art}=0$)[47,49-51], the anomalous Nernst effect (due to a vertical thermal gradient formed by the unbalanced thermal dissipation at the substrate and the surface, $S_{art}\neq0$, $A_{art}=0$)[52], and bulk SOT ($S_{art}\neq0$, $A_{art}\neq0$)[45]. These contributions, if significant, may induce deviation from the linear scaling typically in the thick limit (see data of the Ta/Ni$_{81}$Fe$_{19}$, Ta/Fe, Ta/Fe$_{60}$Co$_{20}$B$_{20}$, and Ta/Fe$_{50}$Pt$_{50}$ devices in Fig. 1d).

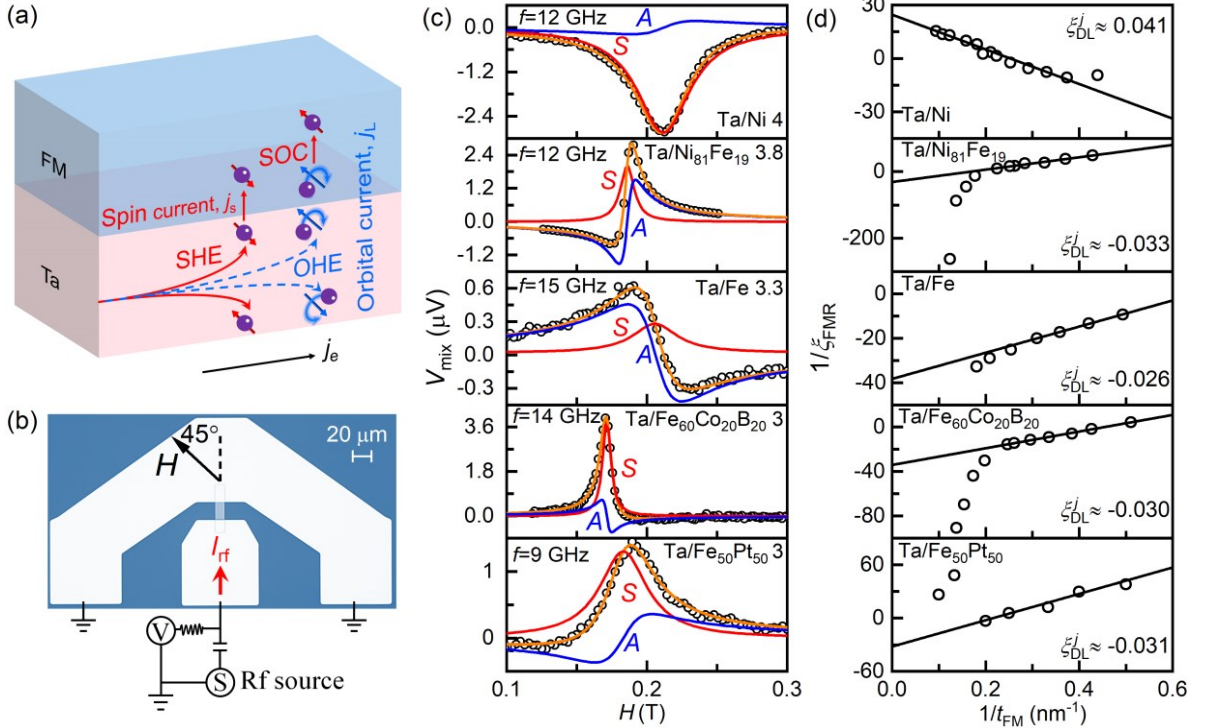

**Fig. 1. ST-FMR analysis.** (a) Schematic of potential torque generation in Ta/ferromagnet bilayer by the spin Hall effect and orbital Hall effect of Ta. (b) Optical microscopy image and measurement geometry of a ST-FMR device. (c) Typical ST-FMR spectra for Ta 5/Ni 4, Ta 5/Ni$_{81}$Fe$_{19}$ 3.8, Ta 5/Fe 3.3, Ta 5/Fe$_{60}$Co$_{20}$B$_{20}$ 3, and Ta 5/Fe$_{50}$Pt$_{50}$ 3 ($\varphi = 45°$), with the three solid curves plotting the best fit of the data to Eq. (1) (in orange), the symmetric (in red), and antisymmetric (in blue) components. (d) Inverse thickness dependence of $1/\xi_{FMR}$ for the Ta/Ni, Ta/Ni$_{81}$Fe$_{19}$, Ta/Fe, Ta/Fe$_{60}$Co$_{20}$B$_{20}$, and Ta/Fe$_{50}$Pt$_{50}$ samples. Solid lines in (d) represent linear fits. The rf power is 8 dBm.

**Absence of orbital current torques.** As summarized in Fig. 2a, the values of $\xi_{DL}^j$ estimated from the above linear-regime analyses remain essentially the same for the Ta/Ni$_{81}$Fe$_{19}$ (-0.033 ± 0.004), the Ta/Fe (-0.026 ± 0.001), the Ta/Fe$_{60}$Co$_{20}$B$_{20}$ (-0.030 ± 0.004), and the Ta/Fe$_{50}$Pt$_{50}$ (-0.031 ± 0.005). This result suggests a constant $\xi_{DL}^j$ for the Ta/FM bilayer with the non-Ni FM thickness within the linear regime (typically <4-5 nm). In contrast, $\xi_{DL}^j$ for the Ta/Ni is estimated to be +0.041 ± 0.003, which is of much higher magnitude and opposite sign compared to others. Such positive estimate for $\xi_{DL}^j$ of Ta/Ni was cited as the key evidence for the presence of a strong orbital current torque in previous reports[19,20,22]. The positive, constant $\xi_{DL}^j$ estimate at the Ni thicknesses of ≥2.5 nm, if represented an orbital current torque, would require the relaxation length of the associated spin and orbital currents to be as short as <0.5 nm (Supplementary Fig. S1), which strongly disagrees with the fairly long relaxation length of nonmagnetic and magnetic metals in some orbital current

torque claims[19,26,29,32-34].

The orbital current torque interpretation of the FM type dependence of the $\xi_{DL}^j$ estimates would require very efficient orbital-spin conversion in the Ni layer and/or at the Ta/Ni interface but negligible orbital-spin conversion in the bulk and interfaces of the non-Ni FMs. According to theories,[42,53] the efficiency of orbital-spin conversion of a bulk material is approximately proportional to the spin-orbit correlation[53], also called spin-orbit polarization[42], due to spin-orbit interaction at the Fermi surface. The spin-orbit correlation is predicted to be linearly correlated to the SOC[42] and to vary in sign with the band filling (positive for the 3d FMs[20] and 5d Pt[42]). While the reliable calculation of the exact values of spin-orbit correlation for the various FMs (i.e., Ni, $Ni_{81}Fe_{19}$, Fe, $Fe_{60}Co_{20}B_{20}$, and $Fe_{50}Pt_{50}$) and the Ta/FM interfaces has remained challenging and beyond the scope of this work, we show below that the bulk and interfacial SOC of the FMs, by themselves, cannot explain the distinct $\xi_{DL}^j$ estimates for the Ta/Ni and Ta/non-Ni FM samples.

In Fig. 2b, we plot the bulk SOC strength of the FMs as 107 meV for Ni, 99 meV for $Ni_{81}Fe_{19}$, 70 meV for Fe, 59 meV for $Fe_{60}Co_{20}B_{20}$, and 335 meV for $Fe_{50}Pt_{50}$, as estimated using the elemental SOC strengths [54]. Here, the average bulk SOC of the Fe alloys ($Ni_{81}Fe_{19}$, $Fe_{60}Co_{20}B_{20}$, and $Fe_{50}Pt_{50}$), which is not trivial to obtain, is estimated assuming a linear composition dependence (see ref. 55 for the case of $KZnSb_{1-x}Bi_x$ alloy) and the composition insensitivity of the elemental SOC of the Fe atoms[56]. The SOC of Ni is only very slightly greater than that of $Ni_{81}Fe_{19}$, Fe, and $Fe_{60}Co_{20}B_{20}$, but 3 times weaker than that of $Fe_{50}Pt_{50}$. The bulk SOC by itself is unlikely to make Ni so different from the other FMs, including $Ni_{81}Fe_{19}$, 81% of which is also Ni. Thus, the FM-type dependence disproves the bulk SOC as the cause of the positive $\xi_{DL}^j$ value for the Ta/Ni. This qualitative conclusion still holds even if the bulk SOC of the Fe alloys ($Ni_{81}Fe_{19}$, $Fe_{60}Co_{20}B_{20}$, and $Fe_{50}Pt_{50}$) was only a monotonic function of the composition (which should be rather reasonable, see ref. 56 for the theoretical results of $Pt_xPd_{1-x}$ alloys) and slightly deviates from the values expected from the linear dependence.

Figure 2c shows the magnetic anisotropy energy density ($K_s$) of the FM interfaces as the indicator of the interfacial SOC[57-59] that is distinct from the bulk SOC and sensitive to the short-range ordering and the spin-orbit proximity effect at the interface. Here, the values of $K_s$ for the Ta/FM samples are determined from the slope of the linear fit of $M_{eff}$ vs $1/t_{FM}$ following the relation of $M_{eff} = M_s - 2K_s/M_st_{FM}$ (Supplementary Fig. S3). Except for the Fe sample with strong SOC at the Fe/MgO interfaces, all the other samples have minimal interfacial SOC at the FM interfaces, revealing that, compared to the non-Ni devices, the Ta/Ni interface does not have a considerably higher interfacial SOC that allowed for enhanced orbital-spin conversion at the Ta/Ni interface and thus enhanced orbital current torque on the Ni layer. Consequently, the positive estimate of $\xi_{DL}^j$ for the Ta/Ni sample cannot be explained by the interfacial SOC. The similar $\xi_{DL}^j$ values of the non-Ni samples suggest that the large $K_s$ of the Ta/Fe sample is mainly from the top Fe/MgO interface because the interfacial SOC of the Ta/Fe interface, if strong, should decrease the spin transparency of the interface and thus $\xi_{DL}^j$ via spin memory loss[8,57].

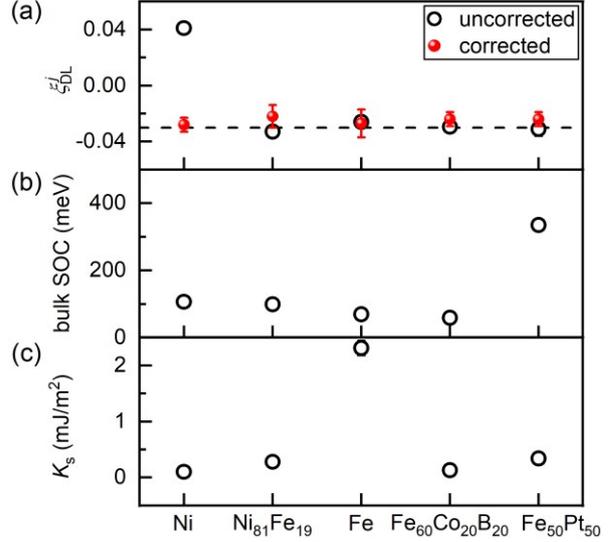

**Fig. 2. Absence of orbital current torque**. (a) Efficiency of the dampinglike spin-orbit torque (black circles for data determined from Fig.1d without correction using control FM single layers, the red dots for data determined from Fig. 3b with correction using the control FM single layers), (b) bulk SOC strength, and (c) Sum interfacial magnetic anisotropy energy density of the magnetic interfaces for the Ta/FM bilayers plotted as a function of the FM type. Error bars are standard deviations.

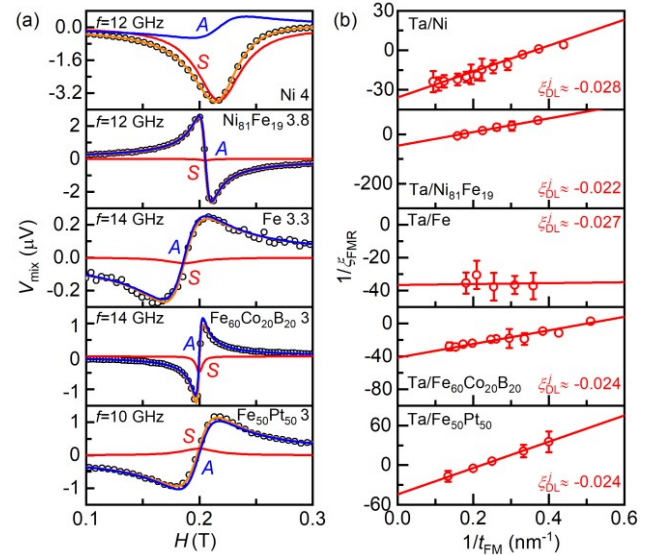

**Fig. 3. Self-induced ST-FMR signal.** (a) Self-induced ST-FMR spectra for Ni 4, $Ni_{81}Fe_{19}$ 3.8, Fe 3.3, $Fe_{60}Co_{20}B_{20}$ 3, and $Fe_{50}Pt_{50}$ 3 single layers ($\varphi = 45°$), with the three solid curves plotting the best fit of the data to Eq. (1) (in orange), the symmetric (in red), and antisymmetric (in blue) components. (b) Inverse thickness dependence of $1/\xi_{FMR}$ contributed from the Ta layers of the Ta/Ni, Ta/$Ni_{81}Fe_{19}$, Ta/Fe, Ta/$Fe_{60}Co_{20}B_{20}$, and Ta/$Fe_{50}Pt_{50}$ devices. $\xi_{FMR}$ is calculated using the $S$ and $A$ values after subtracting the contributions of the magnetic single layers from those measured from the bilayers. Solid lines in (b) represent linear fits. Error bars are standard deviations.

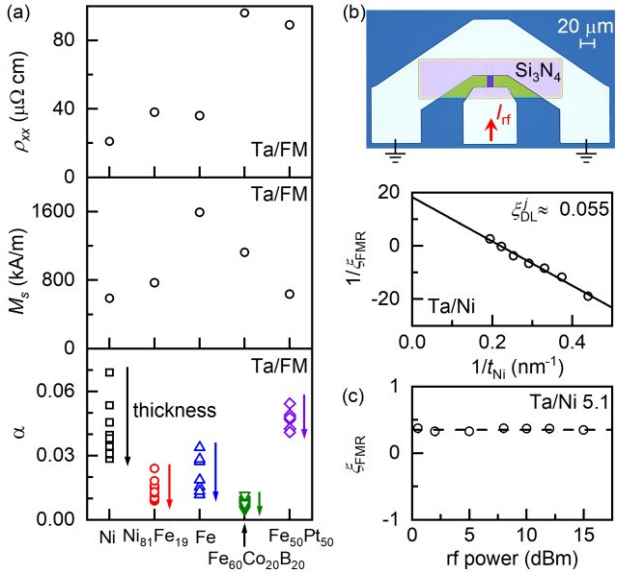

**Fig. 4. Origin of the positive estimate of the SOT for the Ta/Ni devices.** (a) Resistivity, Magnetization, and Magnetic damping of the Ta/FM bilayers with different FM thicknesses. (b) Optical microscopy image and $1/\xi_{FMR}$ vs $1/t_{FM}$ for the Ta/Ni devices with a 150 nm $Si_3N_4$ thermal sink on the top. (c) Minimal variation of $\xi_{FMR}$ with the rf power for Ta 5/Ni 5.1. The solid line in (b) plots the linear fit, while the dashed line in (c) is to guide the eyes.

We discuss below that the positive torque estimate is most likely induced by the presence of a significant thickness-dependent self-induced bulk SOT within the FM layers. Figure 3a shows the representative ST-FMR spectra of the representative FM single layers with the thicknesses at which the Ta/FM bilayers exhibit linear dependence of $1/\xi_{FMR}$ on $1/t_{FM}$ in Fig. 1d. While the self-induced $S$ signals from the non-Ni single-layer FM samples are considerably smaller than that of the bilayers (Fig. 1c), the Ni single layer generates by itself a strong symmetric ST-FMR signal (indicated by the red $S$ curve) that is of the same sign and similar magnitude as that of the Ta/Ni bilayer in Fig. 1c (also see Supplementary Fig. S5 for direct comparison of the measured $S$ and $A$ signals from the control single-layer Ni devices and the Ta/Ni bilayer devices). In consistence with the ST-FMR results, our independent harmonic Hall voltage measurement has also revealed positive bulk SOTs of comparable magnitudes for the Ni and Ta/Ni samples (Supplementary Note 1). These observations consistently suggest the presence of a strong self-induced ST-FMR in the Ni layer in the thickness range of >2 nm. The Ta/$Ni_{81}Fe_{19}$, Ta/Fe, Ta/$Fe_{60}Co_{20}B_{20}$, and Ta/$Fe_{50}Pt_{50}$ samples also demonstrate a deviation from linear scaling only in the thick limit of typically > 4-5 nm (Fig. 1d). These observations are consistent with the previous reports of self-induced bulk SOT in magnetic single layers, e.g., CoPt [45], $Ni_{81}Fe_{19}$ [60], $Fe_{60}Co_{20}B_{20}$ [61], $Fe_{50}Pt_{50}$ [62], and $Fe_xTb_{1-x}$ [63]. The emergence only beyond a non-zero threshold layer thickness is also a characteristic of the bulk SOT effect [45].

After subtracting the self-induced $S$ and $A$ signals of the single FM layers (Fig. 3a) from that of the Ta/FM bilayers (Fig. 1c), the true values of $S$ and $A$ contributed by the Ta layer (see more details in Supplementary Note 2) are used to recalculate $\xi_{FMR}$ of the Ta/FM devices following Eq. (2) and further $\xi_{DL}^j$ of the interfacial dampinglike torque from the linear fits of $1/\xi_{FMR}$ vs $1/t_{FM}$ for the Ta/FM in Fig. 3b following Eq. (3). The linear fit yields a constant $\xi_{DL}^j$ of ≈ -0.028 for the wide Ni thickness range from 2 nm to 10 nm (Fig. 3b), which coincides reasonably with that of non-Ni samples as determined using the same method (the red dots in Fig. 2a) and the harmonic Hall voltage results (Supplementary Fig. S6). These observations are consistent with the negative spin Hall effect of Ta being the dominant source of the spin-orbit torque contribution from the Ta layer. The similar magnitudes of $\xi_{DL}^j$ for different Ta/FM samples are consistent with the previous experiments [58,64] and theories [65] that the spin-mixing conductance of a metallic HM/FM interface is typically robust against the FM type.

Finally, the positive estimate of $\xi_{DL}^j$ in the particularly wide thickness range for the Ta/Ni samples in Fig. 1d cannot be explained by any pumping of spin or orbital current from the FM into the Ta or any anomalous Nernst voltage effect. The contribution of spin/orbital pumping to the $S$ signal via the inverse SHE of Ta should decrease upon enhancement of the conductivity, magnetic damping, and magnetization of the FM (Supplementary Note 3). However, the Ni has the highest conductivity (lowest resistivity) and the highest damping among the studied FMs and similar magnetization as the $Fe_{50}Pt_{50}$ (Fig. 4a and Supplementary Fig. S3). This is consistent with the previous estimation of negligible spin pumping contribution to the ST-FMR signal in the Ta and Pt devices in the pioneering works [1,46]. To test any anomalous Nernst voltage effect, we fabricated ST-FMR devices with a 150 nm $Si_3N_4$ on the top of the Ta/Ni microstrips (Fig. 4b). The thick, high thermal-conductivity [66] $Si_3N_4$ layer is expected to lower or even reverse the rf-heating-induced vertical thermal gradient, if any. However, without the self-torque correction, the $1/\xi_{FMR}$ vs $1/t_{Ni}$ data of these devices still yields a similar positive estimate of $\xi_{DL}^j$ ≈0.055 for the Ta/Ni (Fig. 4b), which is close to that of the devices without any top $Si_3N_4$ sink layer (Fig. 1d). $\xi_{FMR}$ also shows little enhancement as the rf power is increased from 0.01 dBm to 15 dBm (Fig. 4c). The anomalous Nernst effect cannot explain the presence of non-zero $A$ signals in the Ni single layers (Fig. 3a). Thus, the anomalous Nernst voltage is excluded as the cause of the deviation from the linear scaling of $1/\xi_{FMR}$ vs $1/t_{Ni}$ in Fig. 1d and the positive estimate of $\xi_{DL}^j$ for the Ta/Ni samples. These results reveal that the true Ta-contributed dampinglike SOT of the Ta/FM samples is always negative, regardless of the type and the thickness of the FM. This also reaffirms that the orbital Hall effect makes no detectable contribution to the SOT on the FM layer.

**Discussion.** We have established robust evidence for the absence of orbital current torque in the Ta/FM bilayers, regardless of the type and the layer thickness of the FM. These findings have clarified the heated debate over the orbital current torque and have indicated a critical need to recheck the reported orbital current torque in samples containing Ni or other FMs with self-induced ST-FMR signals (whether or not there was a heavy or light metal). The universal deviation from the linear scaling of $1/\xi_{FMR}$ with $1/t_{FM}$ due to self-induced ST-FMR signals suggests that the widely adopted ST-FMR analysis of the dampinglike SOT of a magnetic heterostructure using a single device is highly unreliable for the determination of interfacial SOTs even when the magnetic layer was very thick (> 10 nm) such that the fieldlike SOT contribution became weak. While we do not wish to immediately extrapolate the conclusion of the absence of orbital current torque from the Ta/FM system to other debated material systems, it is important to note that interface-dependent perpendicular effective field[67], FM-type-dependent terahertz emissions,[68,69] and usual magnetoresistance[70] in magnetic heterostructures have been revealed to have physics origins alternative to orbital angular momentum.

**Methods**
**Sample fabrication**. We sputter-deposited in-plane magnetized Ta/FM bilayers with the 5 nm Ta and the FM of Ni, $Ni_{81}Fe_{19}$, $Fe_{60}Co_{20}B_{20}$, Fe, and $Fe_{50}Pt_{50}$ and the control FM single-layer samples with only 1 nm Ta adhesion layer that maintains the Ni property but introduce minimal spin or orbital current. All the samples are grown on oxidized silicon substrates at room temperature with argon pressure of 2 mTorr under base chamber pressure of $10^{-9}$ Torr and protected by a MgO 2nm/Ta 2nm bilayer. These samples are patterned into 10×20 μm$^2$ ST-FMR microstrips and 5×60 μm$^2$ Hall bar devices by photolithography and ion milling. The electrical contacts of each device are Ti 5nm/Pt 150nm as deposited by a sputtering tool. The $Si_3N_4$ thermal sink is also deposited by sputtering.

**Measurement:** The spin-torque ferromagnetic resonance (ST-FMR) measurements were performed using a three-terminal configuration using the longitudinal voltage response by sweeping the in-plane magnetic field (0-0.3 T) at the different azimuth angles relative to the rf current direction. The rf power is modulated at a low frequency for high signal-noise-ratio detection of the ST-FMR response using a lock-in amplifier. The harmonic Hall voltage measurements were performed using a lock-in amplifier to apply an alternating electric field on the Hall bar device and then detect the harmonic Hall voltages while the magnetic field was rotated in the film plane.

**Data availability**
The authors declare that the data supporting the findings of this study are available within the main text and Supplementary Information files.

**Acknowledgments**
This work is supported partly by the Beijing National Natural Science Foundation (Z230006), the National Key Research and Development Program of China (2022YFA1204000), and by the National Natural Science Foundation of China (12304155, 12274405).

**Author contribution**
L.Z. conceived the project, Q.L. performed the measurements, L. Z. and Q. L. wrote the manuscript.

**Competing interests**
The authors declare no competing interest.

**Additional information**
Supplementary information: The online version contains supplementary material available at xxx.

**Correspondence** and requests for materials should be addressed to Lijun Zhu.